\documentclass[12pt]{iopart}
\usepackage{epsfig,graphicx}
\usepackage{iopams}
\begin{document}

\title[Correlated noise effect on semiconductor spin dynamics]
{New insights into electron spin dynamics in the presence of correlated noise}

\author{S Spezia, D Persano Adorno, N Pizzolato and B Spagnolo}

\address{Dipartimento di Fisica, Group of Interdisciplinary Physics\\
Universit\`a di Palermo and CNISM,\\
Viale delle Scienze, edificio 18, I-90128 Palermo, Italy}
\ead{\mailto{stefano.spezia@gmail.com}, \mailto{dominique.persanoadorno@unipa.it}}

\begin{abstract}
The changes of the spin depolarization length in
zinc-blende semiconductors when an external
component of correlated noise is added to a static driving electric
field are analyzed for different values of field strength, noise amplitude and
correlation time. Electron dynamics is simulated by a
Monte Carlo procedure which keeps into account all the possible
scattering phenomena of the hot electrons in the medium and includes
the evolution of spin polarization. Spin depolarization is studied by
examinating the decay of the initial spin polarization of the
conduction electrons through the D'yakonov-Perel process,
the only relevant relaxation mechanism in III-V crystals. Our
results show that, for electric field amplitude lower than the Gunn
field, the dephasing length shortens with the increasing of the noise
intensity. Moreover, a nonmonotonic behavior of spin depolarization
length with the noise correlation time is found, characterized by a maximum
variation for values of noise correlation time comparable with the
dephasing time. Instead, in high field conditions, we find that,
critically depending on the noise correlation time, external
fluctuations can positively affect the relaxation length.
The influence of the inclusion of the electron-electron scattering mechanism is also
shown and discussed.
\end{abstract}


\maketitle

Possible utilization of the electron spin as an information carrier
in electronic devices is an engaging challenge for future spin-based
electronics (spintronics). In particular, in semiconductor
spin-based devices, information is encoded in the electron spin
state, transferred by the conduction electrons, and finally
detected~\cite{FabianDasSarma1999,Awschalom2007,WuReport2010}.
However, a disadvantage of the use of  spin degree of freedom is the
fact that spin polarization is a quantity that is not preserved, as
each initial non-equilibrium magnetization decays both in time and
distance during the transport. Therefore, the spin depolarization
length could be not long enough to permit a reliable manipulation,
control and detection of information.\\
\indent In spintronic devices, the information stored in a system
of polarized electron spins, can be transferred as attached to
mobile carriers by applying an external electric
field~\cite{FabianDasSarma1999,Awschalom2007,WuReport2010,Wolf2001,Zutic2004,Flatte09}.
Because of increasing miniaturization, to avoid too much intense
electric fields, which could lead the system to exhibit a strongly
nonlinear physical behavior, applied voltages are very low. Since,
low voltages are more subjected to the background noise, it becomes
essential to understand the
influence of fluctuations of the electric field on the spin depolarization process.\\
\indent The presence of noise is generally
considered a disturbance, especially in determining the efficiency
of a device, since strong fluctuations can affect
its performance. The existence of fluctuations, for example, can
limit the lifetime of the information stored in a memory cell,
bother the opening (or closure) of random logic gates and cause
the enlargement of the distribution of arrival times of signals in
transmission lines. In quantum
computation, a fundamental problem is the destruction of entangled states of
qubits by interaction with the environment. This event is
characterized by loss of coherence, which is not suitable for the
design of quantum computers~\cite{Palma1996}.\\
\indent In the last decade, however, an increasing interest has been
directed towards possible constructive aspects of noise in the
dynamical response of non-linear systems \cite{Pikovsky,vilar,seol}.
The effect of an external source of noise on the electron transport
in GaAs crystals in the presence of static and/or periodic electric
fields has been studied
~\cite{Varani2005,Persano2008,Persano2009,Persano2011}. Furthermore,
theoretical works which discuss the way to improve the ultra-fast
magnetization dynamics of magnetic spin systems by including random
fields have been recently published
~\cite{Kazantseva2008,Atxitia2009,Bose2010}. Nevertheless, to the
best of our knowledge, the investigation of the role of noise on
the electron spin dynamics in semiconductors is still missing.\\
\indent In this paper, we investigate the spin relaxation process in
low-doped n-type GaAs crystals driven by a randomly fluctuating
electric field. The electron transport is simulated by a Monte Carlo
procedure which takes into account all the possible scattering
phenomena of hot electrons in the medium, and includes the evolution of the spin
polarization vector. The effects caused by the addition of an
external source of correlated noise are investigated by analyzing
the change of the spin depolarization length with respect to the
value obtained in absence of noise. Our findings show that the
presence of a random contribution can affect the value of the
decoherence length, and that the variation is maximum for values of the noise correlation
time comparable with the characteristic time of the spin relaxation
process. Morever, noise induced effects are slightly
reduced by the inclusion of the electron-electron Coulomb interaction.\\

The spin-orbit interaction couples the spin of conduction electrons to
the electron momentum, which is randomized by scattering with impurities
and phonons. The spin-orbit coupling gives rise to a spin precession,
while momentum scattering makes this precession
randomly fluctuating, both in magnitude and orientation~\cite{FabianWu}.\\
\indent For electrons delocalized and under nondegenerate regime,
the D'yakonov-Perel (DP) mechanism \cite{Perel1971,Dyakonov2006} is
the only relevant relaxation process
in n-type III-V semiconductors~\cite{WuReport2010,Jiang09,Litvi2010}.
In a semiclassical formalism, the term of the single
electron Hamiltonian, which accounts for the spin-orbit interaction,
can be written as
\begin{equation}
H_{SO} = \frac{\hbar}{2}{\bf \sigma}\cdot{\bf \Omega}.
\label{HamiltonianSO}
\end{equation}
It represents the energy of electron spins precessing around an
effective magnetic field (${\bf B}=\hbar{\bf \Omega}/\mu_Bg$) with
angular frequency ${\bf \Omega}$, which depends on the orientation
of the electron momentum vector with respect to the crystal axes.
Near the bottom of each valley, the precession vector
can be written as
\begin{equation}
{\bf \Omega}_{\Gamma}=\frac{\beta_{\Gamma}}{\hbar}[k_{x}(k_{y}^{2}-k_{z}^{2})\hat{{\bf x}}+k_{y}
(k_{z}^{2}-k_{x}^{2})\hat{{\bf y}}+k_{z}(k_{x}^{2}-k_{y}^{2})\hat{{\bf z}}]
\label{effectivefieldgammavalley}
\end{equation}
in the $\Gamma$-valley~\cite{Dresselhaus55},
\begin{equation}
{\bf \Omega}_{L}=\frac{\beta_{L}}{\sqrt{3}}[(k_{y}-k_{z})\hat{{\bf x}}+(k_{z}-k_{x})\hat{{\bf y}}+(k_{x}-k_{y})\hat{{\bf z}}]
\label{effectivefieldLvalley}
\end{equation}
in the L-valleys located along the [111] direction in the crystallographic axes, and
\begin{equation}
{\bf \Omega}_{X}=\beta_{X}[-k_{y}\hat{{\bf y}}+k_{z}\hat{{\bf z}}]
\label{effectivefieldXvalley}
\end{equation}
in the X-valleys located along the [100]
direction~\cite{Saikin2006}. In equations
(\ref{effectivefieldgammavalley})-(\ref{effectivefieldXvalley}),
$k_{i}$ ($i=x,y,z$) are the components of the electron wave vector.
$\beta_{\Gamma}$, $\beta_{L}$ and $\beta_{X}$ are the spin-orbit
coupling coefficients. Here, we assume $\beta_{L}$=$0.26$
eV$/${\AA}$\cdot2/\hbar$ and $\beta_{X}$=$0.059$
eV$/${\AA}$\cdot2/\hbar$, as recently theoretically estimated
in Ref.~\cite{Fu2008}, while $\beta_{\Gamma}$
is calculated as in Ref.~\cite{Spezia2010_2}.\\
\indent Since the quantum-mechanical description of the electron
spin evolution is equivalent to that of a classical momentum ${\bf
S}$ experiencing the magnetic field ${\bf B}$, we
describe the spin dynamics by the classical equation of precession
motion
\begin{equation}
\frac{d{\bf S}}{dt}={\bf \Omega}\times{\bf S}. \label{Poisson}
\end{equation}
\indent The DP mechanism acts between two scattering events and
reorients the direction of the precession axis and of the effective
magnetic field ${\bf B}$ in random and trajectory-dependent way.
This effect leads the spin precession frequencies ${\bf \Omega}$ and
their directions to vary from place to place within the electron spin
ensemble. This spatial variation is called {\it inhomogeneous
broadening}~\cite{Slichter}.\\

When a semiconductor is in contact with a spin polarization source
at $x=0$, in the absence of any magnetic field (${\bf B}={\bf 0}$)
and with a static electric field directed along the $\hat{{\bf
x}}$-direction  (${\bf F}=F\hat{{\bf x}}$), the drift-diffusion
model predicts an exponential decay of the spin polarization
$S(x)=S_0\exp(-x/L)$, in which $L$ is the electric-field-dependent
spin depolarization length~\cite{Yu2002}. The electron spin
dephasing length can be seen as the distance that the electrons move
with average drift velocity $v_d$ within the spin lifetime $\tau$
\cite{Yu2002},
\begin{equation}
L=v_d \tau=\mu F \tau=\frac{qF \tau_p \tau}{m^*},
\label{length}
\end{equation}
where $\mu=q\tau_p/m^*$ is the electron mobility, $q$ the elementary
charge, $\tau_p$ the momentum relaxation time and $m^*$ the electron
effective mass.\\
\indent The inhomogeneous broadening, quantified by the average
squared precession frequency $<\mid{\bf \Omega}({\bf k})\mid^2>$,
together with the correlation time of the random angular diffusion
of spin vector $\tau_c$ are the relevant variables in the D'yakonov-Perel's
formula
\begin{equation}
\tau^{-1}=<\mid{\bf \Omega}({\bf k})\mid^2>\tau_c.
\label{DPformulae}
\end{equation}
Since, in our case, we include the electron-electron interaction
mechanism, one needs to distinguish between the momentum relaxation
time $\tau_p$ and the momentum redistribution time $\tau^{'}_p$,
which is practically equal to $\tau_c$. This distinction is
necessary because, although electron-electron scattering contributes
to momentum redistribution, it
does not directly lead to momentum relaxation~\cite{Glazov2002,Kamra2011}.\\
\indent By using equations (\ref{length}) and (\ref{DPformulae}),
the spin depolarization length can be expressed in the form,
\begin{equation}
L=\frac{eF}{m^*}\frac{1}{<\mid{\bf \Omega}({\bf k})\mid^2>}\frac{\tau_p}{\tau^{'}_p}.
\label{Ld}
\end{equation}

In our simulations the semiconductor bulk is driven by a fluctuating
electric field
\begin{equation}
F(t)=F_0+\eta(t) \label{electricfield}
\end{equation}
where $F_0$ is the amplitude of the deterministic part and $\eta(t)$
is a random term, modeled by a stochastic process. Here, $\eta(t)$
is modeled as an Ornstein-Uhlenbeck (OU) process, which obeys the
stochastic differential equation ~\cite{Gardiner1997}:
\begin{equation}
\frac{\textrm{d}\eta(t)}{\textrm{d}t}=-\frac{\eta(t)}{\tau_D}+\sqrt{\frac{2D}{\tau_D}}\xi(t)
\label{Ornstein}
\end{equation}
where $\tau_D$ and $D$ are, respectively, the correlation time and
the intensity of the noise described by the OU process with
autocorrelation function\\
$\langle\eta(t)\eta(t')\rangle=D\exp(-|t-t'|/\tau_D)$. $\xi(t)$ is
a Gaussian white noise with zero mean $<\xi(t)>=0$, and autocorrelation function
$\langle \xi(t)\xi(t')\rangle=\delta(t-t')$.
Within the framework of the Ito's calculus, the general solution of the
equation~(\ref{Ornstein}) leads to the following expression for the
stochastic evolution of the amplitude of the electric field
\begin{equation}
F(t)=F_0+\eta(0)e^{-t/\tau_D}+\sqrt\frac{2D}{\tau_D}\int_0^te^{-\frac{t-t'}{\tau_D}}dW(t'),
\label{OUsol}
\end{equation}
where the initial condition is $\eta(0)=0$, and $W(t)$ is the Wiener process~\cite{Gardiner1997}.\\
\indent In a practical system, $\eta(t)$ could be generated by a RC
circuit driven by a source of Gaussian white noise, with correlation
time $\tau_D=(RC)^{-1}$ (see equation (\ref{Ornstein})). The
Gaussian white noise can be generated either by the Zener breakdown
phenomenon in a diode or in a inversely polarized base-collector
junction of a BJT, either by amplifying the thermal noise in a
resistor \cite{Askari2010}. The correlation time $\tau_D$ is tunable
by using a diode (varicap) with a voltage-dependent variable
capacitance; the noise intensity $D$ can be chosen, for example, by
suitably amplifying the noise produced through the Zener stochastic
process.

 \begin{figure} [htbp]
 \begin{center}
 \resizebox{0.9\columnwidth}{!}{%
 \includegraphics{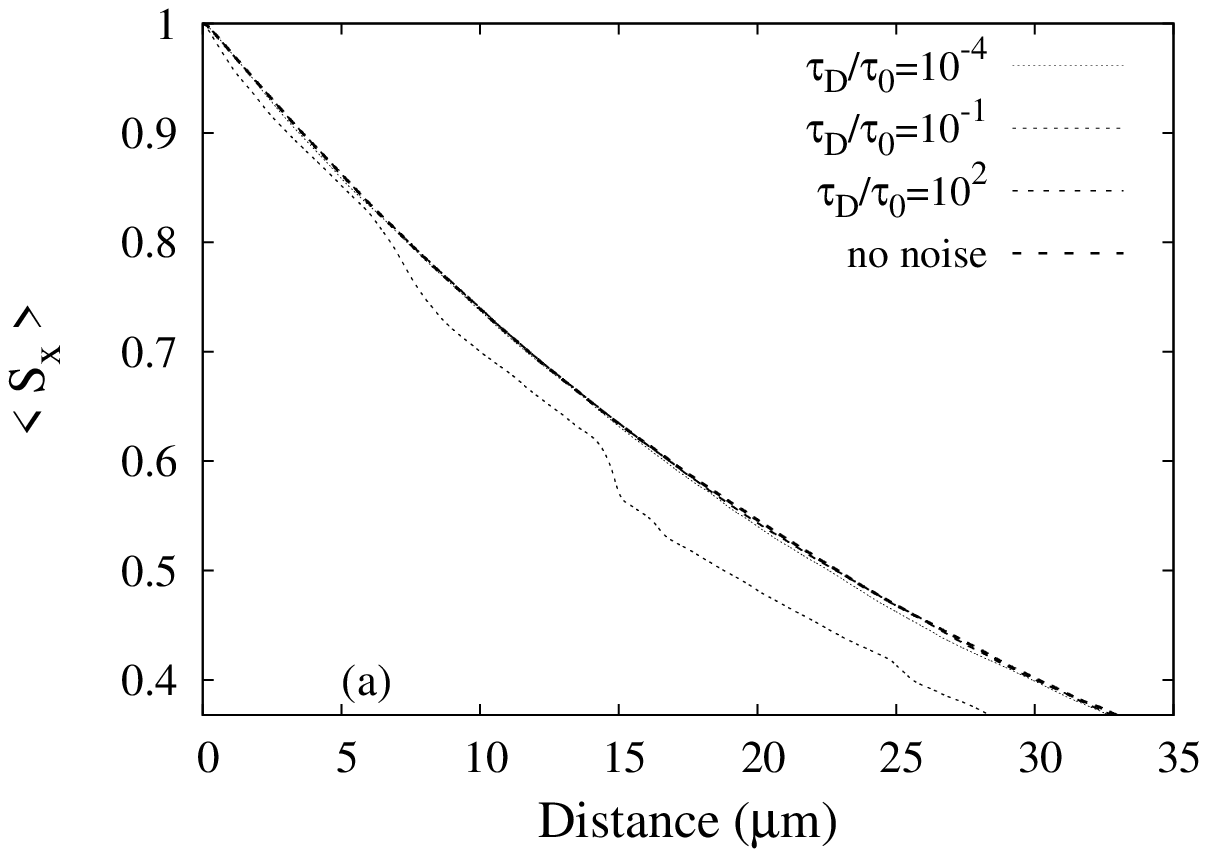}
 \includegraphics{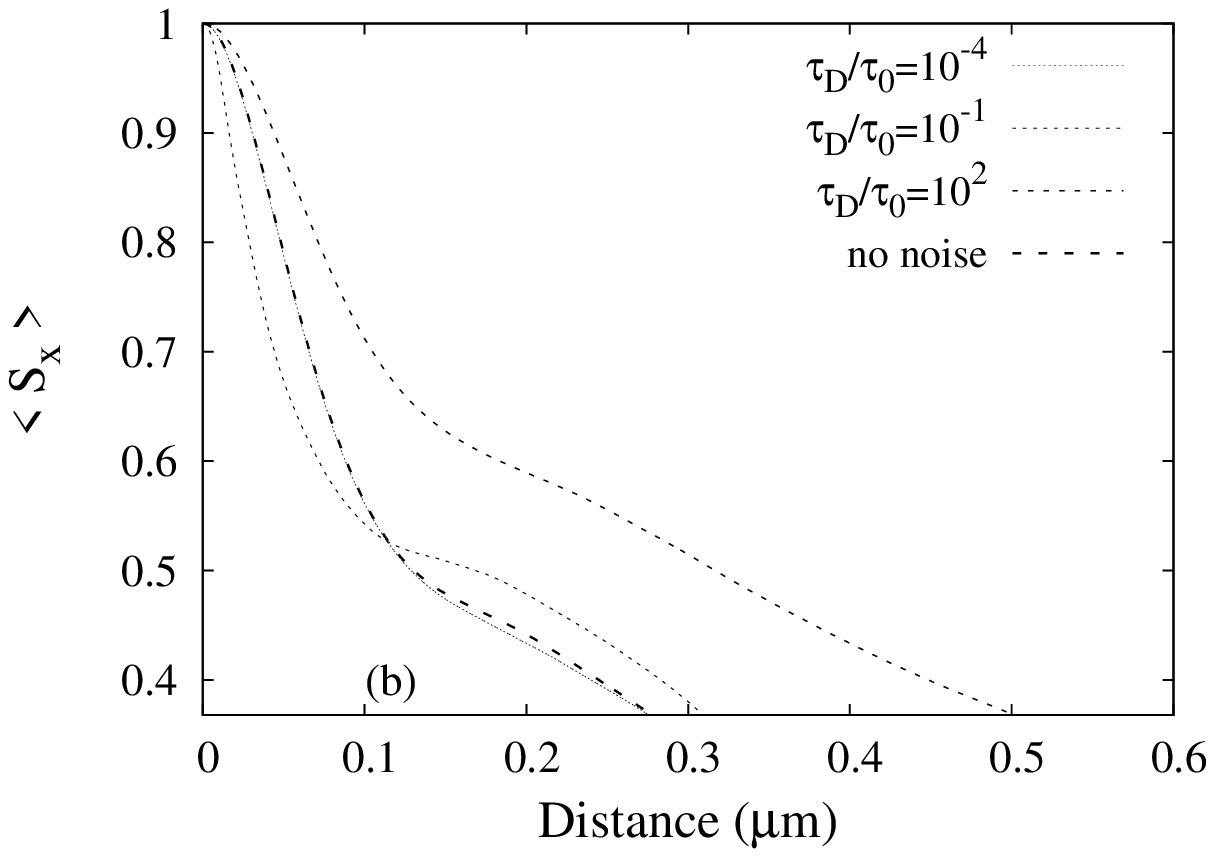}
 }
 \caption{Spin polarization $\langle S_{x}\rangle$ as a
function of the distance from the injection point ($x=0$) obtained
by applying a fluctuating field characterized by a deterministic
component $F_0$ and a random component with standard deviation
$D^{1/2}$, for different values of noise correlation time $\tau_D$:
$10^{-4}\tau_0$, $10^{-1}\tau_0$, $10^2\tau_0$, and in absence of
noise. (a) $F_0=1$ kV/cm, $D^{1/2}=0.6$ kV/cm and (b) $F_0=6$
kV/cm, $D^{1/2}=3.6$ kV/cm}
 \label{fig:1}       
 \end{center}
 \end{figure}

\noindent The Monte Carlo code used here follows the procedure
described in Ref.~\cite{Persano2000}. The spin polarization vector
has been incorporated into the algorithm as an additional quantity
and calculated for each free carrier. Furthermore, the
electron-electron scattering is included, as extensively described
in Ref.~\cite{SpeziaNEW}. All simulations are performed in a GaAs
crystal with a free electrons concentration equal to $10^{13}$
cm$^{-3}$ and lattice temperature $T_L$ equal to $77$ K. The
temporal step is $10$ fs and an ensemble of $5\cdot10^{4}$ electrons
is used to collect spin statistics. Moreover, we assume that all
donors are ionized and that the free electron concentration is equal
to the doping concentration. All physical quantities of interest are
calculated after a transient time of typically $10^{4}$ time steps,
long enough to achieve the steady-state transport regime. The spin
relaxation simulation starts with all electrons of the ensemble
initially polarized $(\langle {\bf S}\rangle={\bf 1})$ along
$\hat{{\bf x}}$-axis of the crystal, at the injection plane
$(x_{0}=0)$~\cite{Zutic2004,WuReport2010}. The spin lifetime $\tau$
and the spin depolarization length $L$ are calculated by extracting,
respectively, the time and the distance from the injection plane of
the center of mass of the electron ensemble, corresponding to a
reduction of the initial spin polarization by a factor $1/e$.
 \begin{figure} [htbp]
 \begin{center}
 \resizebox{0.6\columnwidth}{!}{%
 \includegraphics{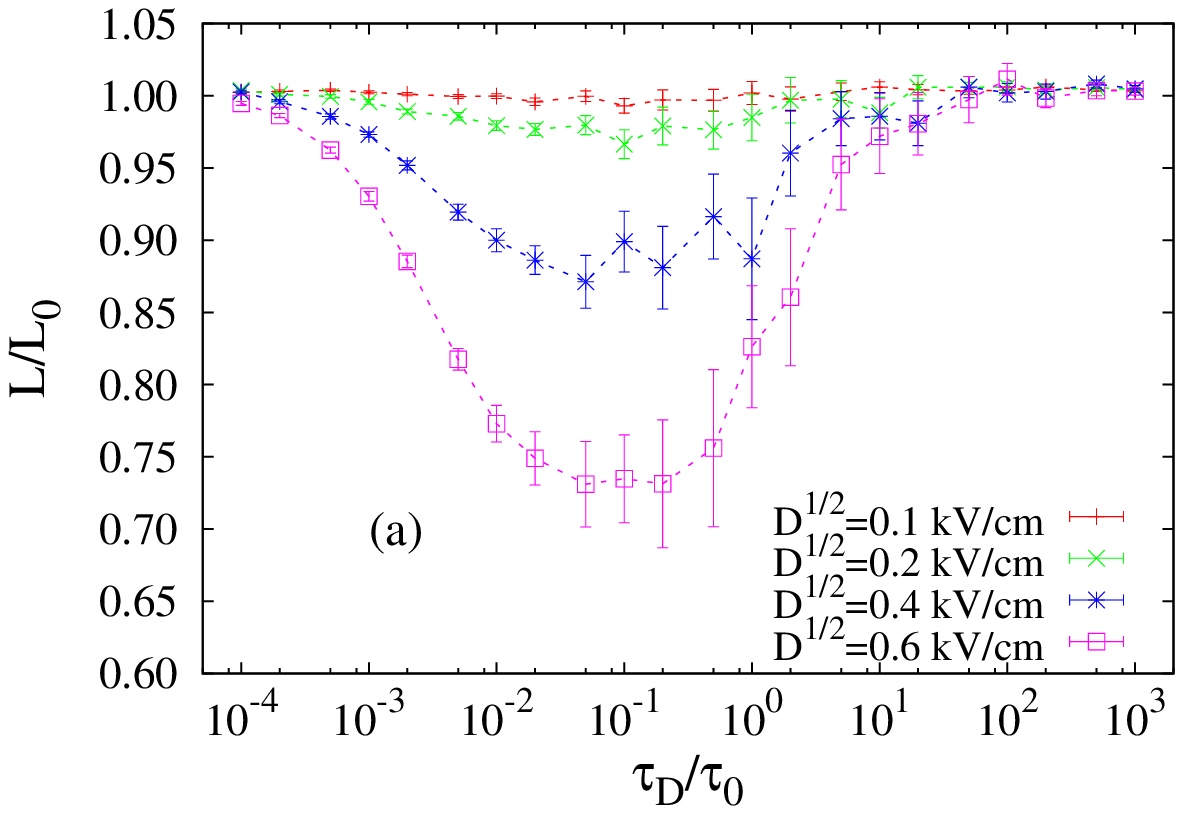}
 }
 \resizebox{0.9\columnwidth}{!}{%
 \includegraphics{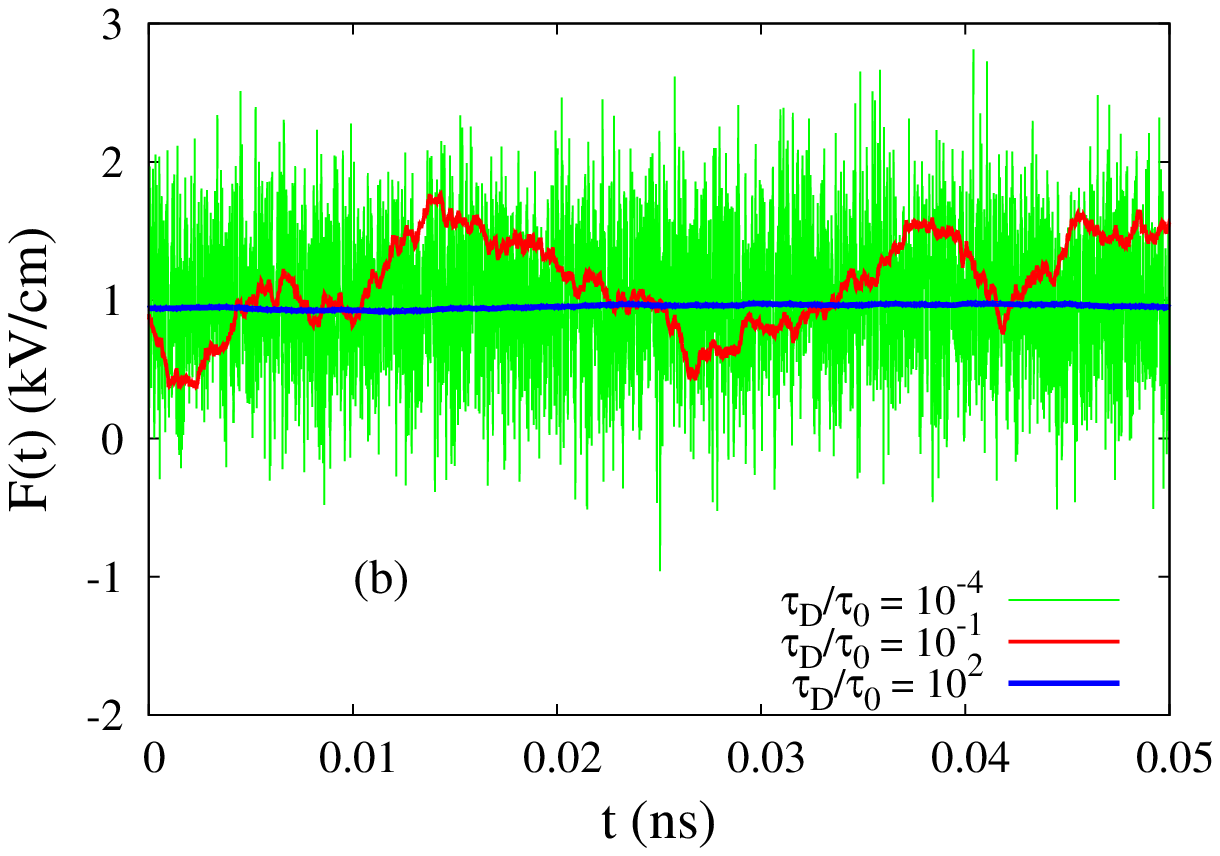}
 \includegraphics{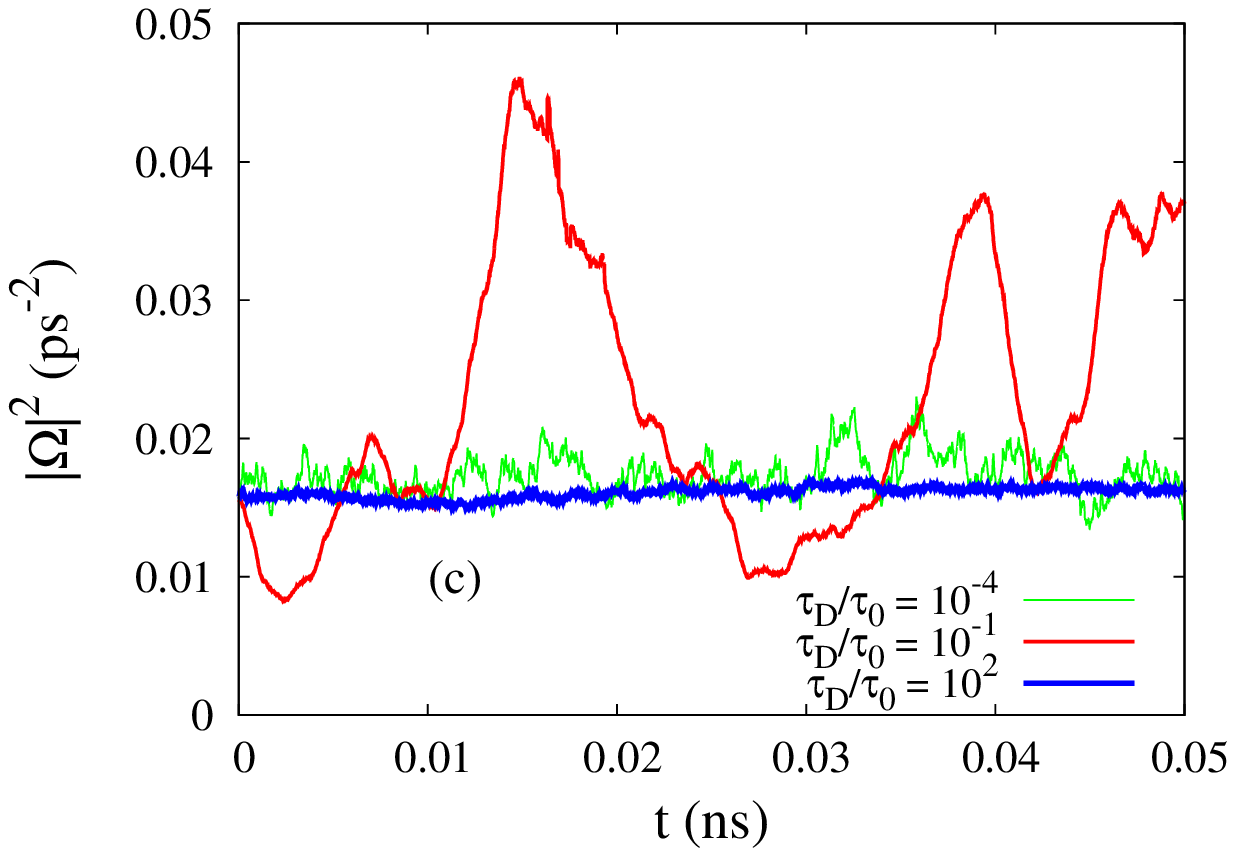}
 }

 \caption{(a) Ratio between the spin depolarization length $L$
 in the presence of noise and $L_0$, obtained in absence of noise,
 as a function of the ratio between the noise correlation time $\tau_D$ and the spin
 relaxation time in absence of noise $\tau_0$, at different
values of noise intensity $D$. (b) Electric field amplitude $F(t)$
and (c) squared precession frequency $\mid{\bf \Omega}({\bf
k})\mid^2$ as a function of time $t$. $F_0=1$ kV/cm.}
 \label{fig:2}       
 \end{center}
 \end{figure}

In Figure~\ref{fig:1}, we show the electron spin average
polarization $\langle S_{x}\rangle$ as a function of the distance
traveled by the center of mass of the electron cloud from the
injection point, in the presence of a fluctuating field
characterized by a deterministic component with amplitude $F_0$ and
a random component with standard deviation $D^{1/2}$, for three
different values of noise correlation time $\tau_D$:
$10^{-4}\tau_0$, $10^{-1}\tau_0$, $10^2\tau_0$ and in absence of
noise; $\tau_0$ is the spin
 relaxation time obtained in absence of noise at the same value of $F_0$.
In panel (a): $F_0=1$ kV/cm and $D^{1/2}=0.6$ kV/cm; in panel (b): $F_0=6$ kV/cm
and $D^{1/2}=3.6$ kV/cm.\\
\indent When $\tau_D<<\tau_0$, the
spin dephasing process is not affected by the fluctuations of the
electric field, which have a negligible memory ($\tau_D$) with
respect to the characteristic time $\tau_0$ of the system. In
particular, under low electric field condition [panel (a)], the spin
relaxation process is significantly influenced by the fluctuating
field only for values of noise correlation time comparable with
$\tau_0$, while the process becomes
quasi-deterministic when $\tau_D>>\tau_0$. Differently, under high electric field
condition [panel (b)] when $\tau_D\sim\tau_0$ and $\tau_D>>\tau_0$, a
slow down of the spin relaxation process is observed.\\
\indent With the aim of investigating the effects of the correlated
noise source on the spin depolarization process, we performed $100$
different realizations and evaluated both average values and error
bars of the extracted spin depolarization lengths.\\
\indent In panel (a) of Figure~\ref{fig:2}, we show the ratio
between the spin depolarization length $L$ in the presence of noise
and $L_0$, obtained in absence of noise, as a function of the ratio
between the noise correlation time $\tau_D$ and $\tau_0$, at
different values of noise intensity $D$ and with $F_0=1$ kV/cm. For
these values of parameters, $L_0=32.6$ $\mu$m and $\tau_0=0.16$ ns.
The addition of a source of correlated fluctuations, characterized
by $10^{-2}\tau_0<\tau_D<\tau_0$, reduces the values of the spin
depolarization length $L$ up to 25\%. In particular, $L/L_0$ is a
nonmonotonic function of $\tau_D/\tau_0$ which exhibits a minimum
for $\tau_D/\tau_0\approx0.1$. For both $\tau_D<<\tau_0$ and
$\tau_D>>\tau_0$, the values of $L$ coincide with those of $L_0$.
The presence of the minimum, which becomes deeper with the
increasing of the noise amplitude, can be explained by analyzing the
temporal evolution of the quantities related to the electron
transport and to the spin relaxation process. In panels (b) and (c)
of Figure~\ref{fig:2}, we report the electric field amplitude $F(t)$
and the squared precession frequency $\mid{\bf \Omega}({\bf
k})\mid^2$, respectively, as a function of time $t$. For very low
values of $\tau_D/\tau_0$, $\mid{\bf \Omega}({\bf k})\mid^2$
symmetrically fluctuates around its average value, corresponding to
that obtained in absence of noise. By increasing the value of
$\tau_D$, the effective electric field felt by electrons, within a
time window comparable with the spin relaxation time, becomes very
different from the value $F_0$ [panel (b)]. As a consequence, the
temporal evolution of $\mid{\bf \Omega}({\bf k})\mid^2$ shows an
evident asymmetry in the same temporal window [panel (c)]. Because
of the proportionality between the electron momentum $k_{x}$ and the
electric field $F(t)$, the equation
(\ref{effectivefieldgammavalley}) leads to a quadratic relation
between $F(t)$ and $\mid{\bf \Omega}({\bf k})\mid^2$ on the
$k_{x}^2(k_{y}^{2}-k_{z}^{2})^2$ term and at fourth power on the
other two terms. Hence, the values of $F$ greater than $F_0$ give
rise to values of $\mid{\bf \Omega}({\bf k})\mid^2$ much greater
than those obtained for $F<F_0$. So, in accordance with
equation~(\ref{DPformulae}), the asymmetry of $\mid{\bf \Omega}({\bf
k})\mid^2$ is responsible for the observed reduction of spin
lifetime. By further increasing the value of $\tau_D$, the random
fluctuating term $\eta(t)$ of the electric field tends to its
initial value $\eta(0)=0$ [see equation~(\ref{Ornstein})], and
$F(t)\rightarrow F_0$. Therefore, the behavior of system becomes
quasi-deterministic and the spin dephasing length $L$ approaches
its deterministic value $L_0$.\\

 \begin{figure} [htbp]
 \begin{center}
 \resizebox{0.90\columnwidth}{!}{%
 \includegraphics{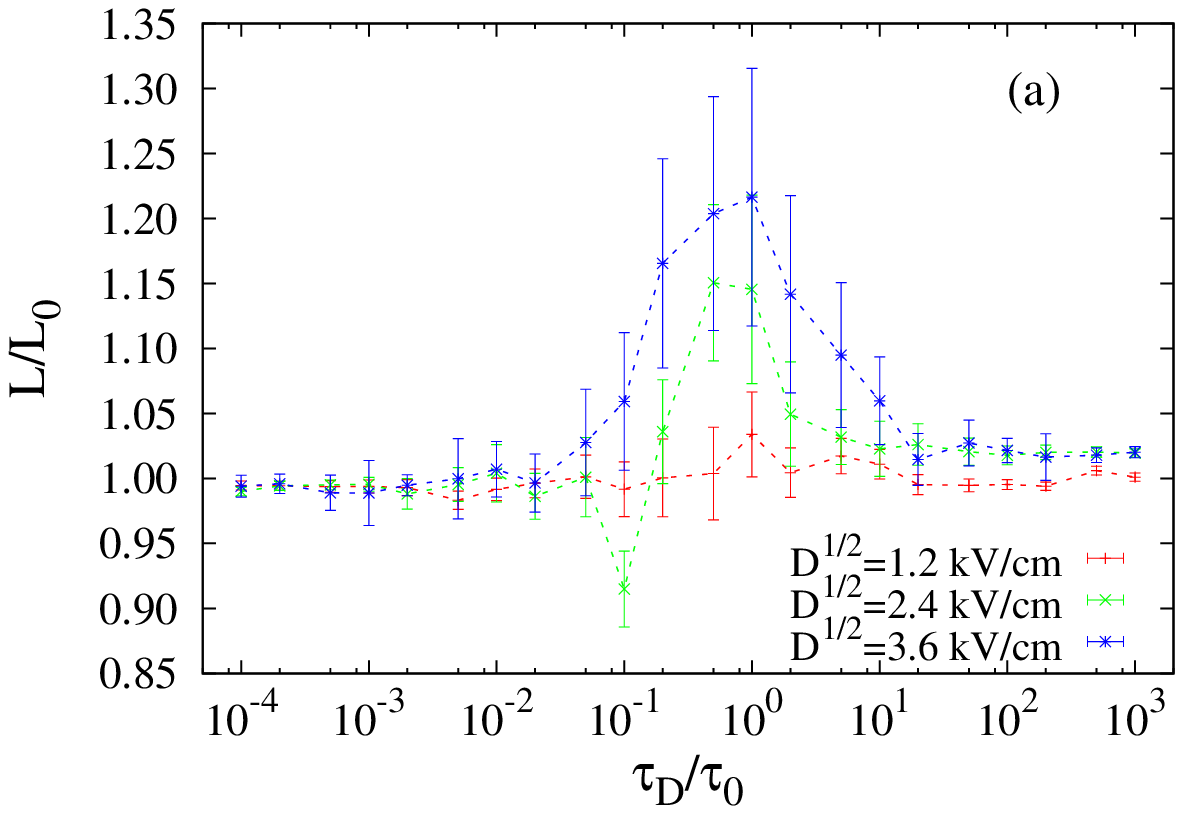}
 \includegraphics{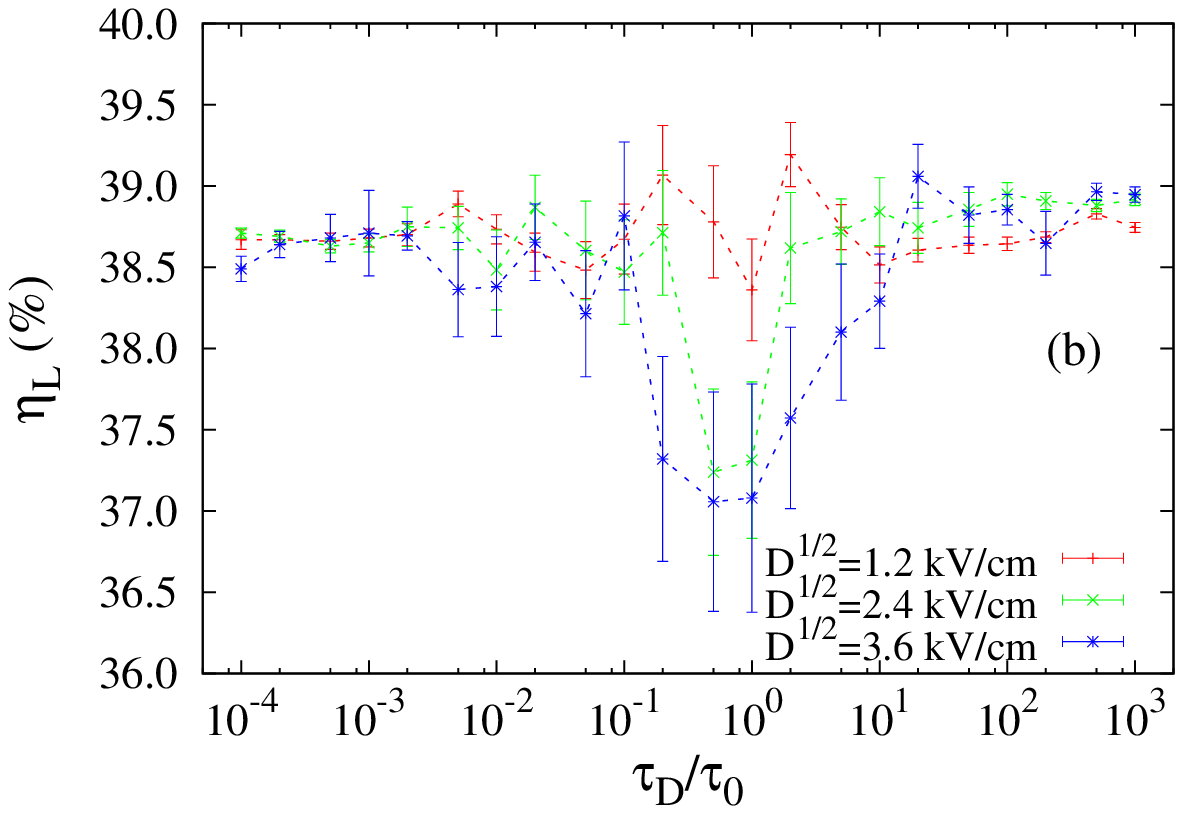}
 }
 \caption{(a) Ratio between the spin depolarization length $L$
 in the presence of noise and $L_0$, obtained in absence of noise,
 as a function of the ratio between the noise correlation time $\tau_D$ and the spin
 relaxation time in absence of noise $\tau_0$ and (b) electron occupation
percentage in L-valleys $\eta_L$ as a function of $\tau_D/\tau_0$,
at different values of noise intensity $D$. $F_0=6$ kV/cm.}
 \label{fig:3}       
 \end{center}
 \end{figure}

\indent In Figure~\ref{fig:3} (a), we show the ratio $L/L_0$ as a
function of $\tau_D/\tau_0$ for $F_0=6$ kV/cm, at different values
of noise intensity $D$. For these values of parameters, $L_0=279$ nm
and $\tau_0=1.13$ ps. In the presence of a driving electric field
greater than the necessary static field to allow the electrons to
move towards the upper energy valleys, the Gunn field $E_G=3.25$
kV/cm, we find a positive effect of the field fluctuations. In fact,
despite the error bars are large, our findings show that the
addition of correlated fluctuations, characterized by
$10^{-1}\tau_0<\tau_D<\tau_0$, can increase the value of the spin
depolarization length $L$ up to the 20\% of $L_0$. This effect is
maximum for $\tau_D/\tau_0\approx1$. For the reasons discussed
above, even in the high field case, both for very high and very low
values of noise correlation time $\tau_D$, the value of $L$
approaches $L_0$. The presence of a positive effect of noise can be
ascribed to the reduction of the electron occupation percentage in
$L$-valleys, shown in panel (b) of Figure~\ref{fig:3}. This finding,
which can be considered as a further example of Noise Enhanced
Stability (NES)~\cite{Persano2009,Agudov2001,Fiasconaro2009}, leads a greater number of
electrons to experience a spin-orbit coupling in $\Gamma$-valley at
least of one order of magnitude weaker than that present in
$L$-valleys ~\cite{Spezia2010_2}, causing a decrease of efficacy of
the DP dephasing mechanism.

 \begin{figure} [htbp]
 \begin{center}
\resizebox{0.90\columnwidth}{!}{%
 \includegraphics{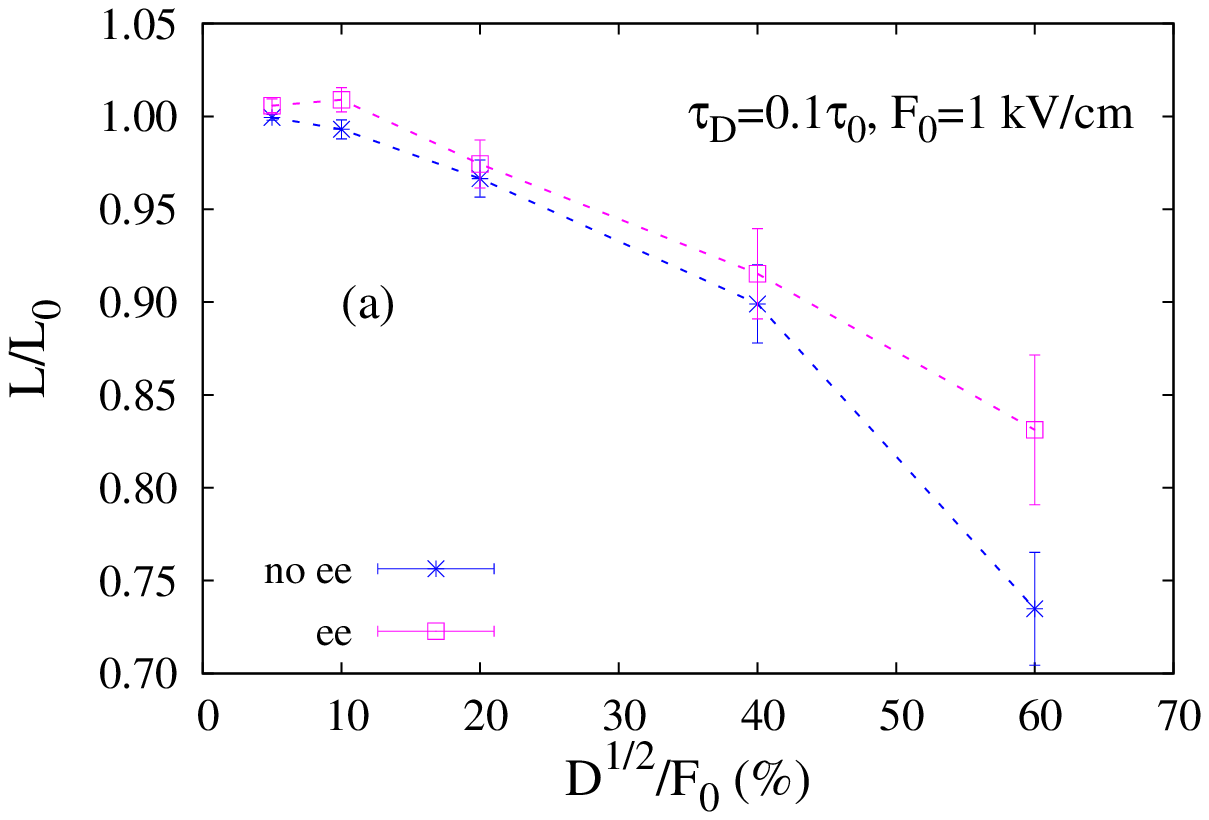}
 \includegraphics{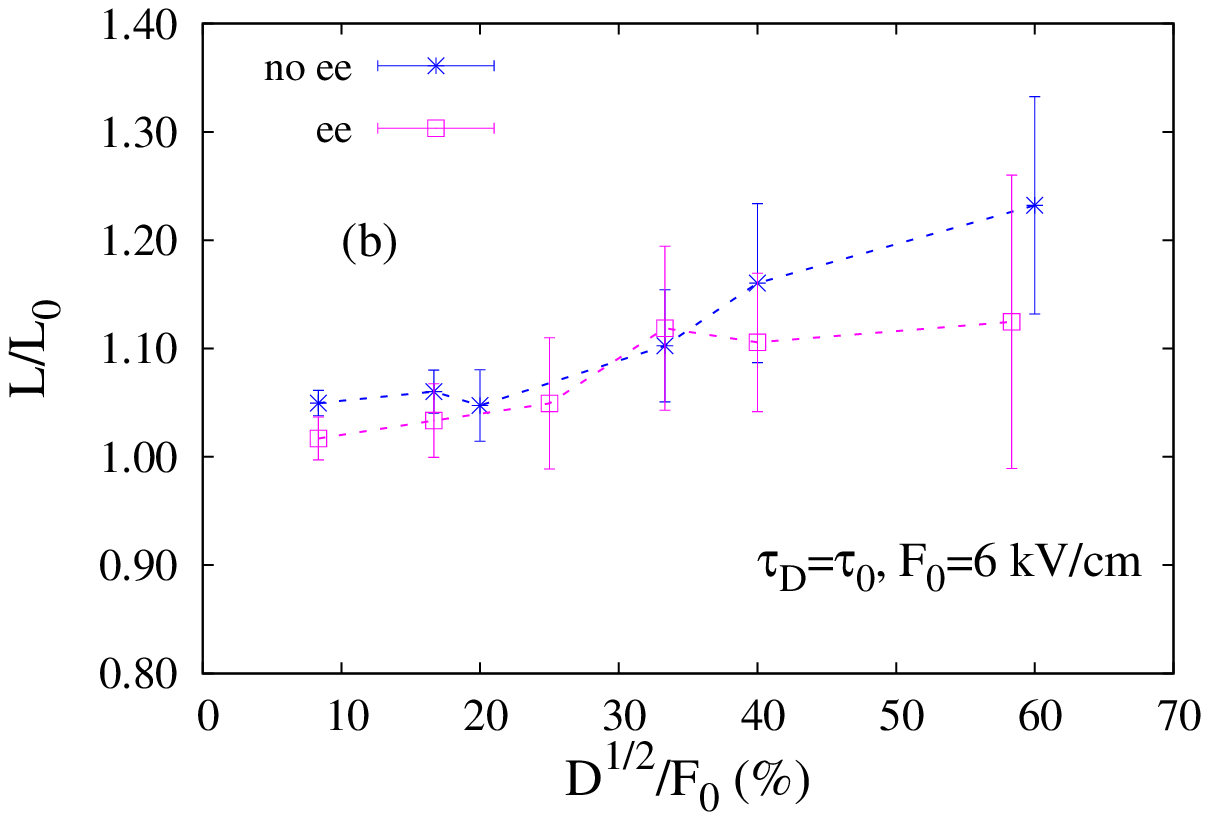}
 }
 \caption{Ratio between the spin depolarization length $L$
 in the presence of noise and $L_0$, obtained in absence of noise,
 as a function of the ratio between the noise amplitude $D^{1/2}$ and $F_0$.
The ee-points are obtained through full calculations including the
electron-electron scattering mechanism; the no ee-points are
calculated without it. (a)
$\tau_D=0.1\tau_0$ and $F_0=1$ kV/cm, (b) $\tau_D=\tau_0$ and
$F_0=6$ kV/cm.}
 \label{fig:4}       
 \end{center}
 \end{figure}

Although for many decades it has been believed that the
electron-electron (e-e) Coulomb scattering does not contribute to
the spin relaxation/dephasing process, in the presence of
inhomogeneous broadening in spin precession, each kind of
scattering, including spin-conserving scattering, can cause
irreversible spin dephasing. Moreover, the Coulomb interaction is
fundamental to obtain the correct distribution of the electrons
in the ${\bf k}$-space~\cite{WuReport2010,Kamra2011}.
Recently, it has been  shown that e-e scattering inclusion
leads to a strong increase of spin relaxation time in bulk
semiconductors, in a wide range of electric field amplitude,
lattice temperature and doping density~\cite{SpeziaNEW}.\\
\indent In order to quantify the effect of the Coulomb interaction
on the spin depolarization process in the presence of external
noise, we show a comparison between the results obtained with the
e-e scattering mechanism and without it, analyzing the behavior of
$L/L_0$, as a function of the ratio between the noise standard
deviation $D^{1/2}$ and the deterministic value of the driving field
$F_0$ (see Figure~\ref{fig:4}). We show the comparison at the values
of noise correlation time corresponding with the maximum of the
noise effect observed by neglecting the Coulomb interaction. Each
panel of Figure~\ref{fig:4} shows the "ee-points", obtained through
full calculations including the electron-electron scattering
mechanism, and the "no ee-points", calculated without the
electron-electron interaction. In panel (a): $\tau_D=0.1\tau_0$ and
$F_0=1$ kV/cm, with $L_0=127$ $\mu$m and $\tau_0=0.49$ ns; in panel
(b): $\tau_D=\tau_0$ and $F_0=6$ kV/cm, with $L_0=666$ nm and
$\tau_0=2.42 $ ps; the new values of $\tau_0$ and $L_0$ have been
obtained by including the Coulomb scattering. For
$\tau_D=0.1\tau_0$, the reduction of the spin depolarization length,
caused by fluctuations, is slightly affected from the inclusion of
electron-electron scattering mechanism up to values of $D^{1/2}$
lower than 40\% of the value of $F_0$. For values of $D^{1/2}/F_0$
greater than 0.4, the inclusion of the e-e scattering leads to a
longer spin dephasing length, i.e. the electron-electron scattering
mechanism reduces the negative effect of noise. This effect of the
Coulomb interaction could be ascribed to the frequent momentum
redistribution experienced from the electrons
ensemble~\cite{WuReport2010}. Unfortunately, the Coulomb interaction
inclusion seems to randomize the system also for field amplitude
greater than the Gunn field. Under high electric field conditions,
in fact, up to values of $D^{1/2}$ lower than 33\% of the value of
$F_0$, the quantity $L/L_0$ is almost not influenced by the
inclusion of the e-e mechanism. By increasing the noise amplitude, a
slight noise-induced positive effect on spin relaxation length is
found. In this case, the addition of a source of correlated
fluctuations, having correlation time comparable with the spin
lifetime, enhances the value of the spin depolarization length $L$
of only about 10-15\%.

In this work, for the first time, we have investigated the noise
influence on the electron spin relaxation process in lightly $n$-doped
GaAs semiconductor bulks by also including the electron-electron
interaction. The findings show that a fluctuating electric field,
obtained by adding a correlated source of noise to a static field,
can modify the spin depolarization length. For electric fields lower
than the Gunn field and values of the noise correlation time
$\tau_D\sim\tau_0$, the spin lifetime obtained in absence of noise,
a reduction of the spin depolarization length up to 15\% has been
observed, strongly dependent on the noise intensity. This behavior
can be explained by the different effective electric field
experienced by the electron ensemble, within a time window
comparable with $\tau_0$. On the contrary, in the high electric
field regime, for $\tau_D$=$\tau_0$, we find an enhancement of the
spin relaxation length up to 15\%. This positive effect can be
explained by the decrease of the occupation of the $L$-valleys,
where the strength of spin-orbit coupling felt by electrons is at
least one order of magnitude greater than that present in $\Gamma$-valley and represents
an example of NES in spin depolarization process.\\
\indent To conclude, our preliminary results show that the presence
of fluctuations in applied voltages changes the maintenance of long
spin depolarization lengths in a way strongly dependent  on both the
strength of the applied electric field and the noise correlation
time. Further studies are needed to learn more about the relationship
between the semiconductor characteristic time scales, the noise
correlation time and the e-e interaction, in order to
find the most favorable conditions for the manipulation of electron
spins.

\ack This work was partially supported by MIUR and CNISM. The
authors acknowledge CASPUR for the computing support via the
standard HPC grant 2010.

\end{document}